\documentclass[prb,twocolumn,amssymb,showpacs]{revtex4}

\usepackage{bm}
\usepackage{graphicx}
\usepackage{dcolumn}
\begin{document}

\title{Polariton lasing in a multilevel quantum dot \\
strongly coupled to a single photon mode}
\author{Carlos Andr\'es Vera$^{(a)}$, Herbert Vinck-Posada$^{(a)}$, 
 and Augusto Gonz\'alez$^{(b)}$}
\affiliation{$^{(a)}$Instituto de F\'{\i}sica, Universidad de Antioquia,
 AA 1226, Medell\'{\i}n, Colombia\\
 $^{(b)}$Instituto de Cibern\'etica, Matem\'atica y 
 F\'{\i}sica, Calle E 309, Vedado, Ciudad Habana, Cuba}

\begin{abstract}
We present an approximate analytic expression for the photoluminescence
spectral function of a model polariton system, which describes a quantum
dot, with a finite number of fermionic levels, strongly interacting with
the lowest photon mode of a pillar microcavity. Energy eigenvalues and 
wavefunctions of the electron-hole-photon system are obtained by numerically
diagonalizing the Hamiltonian. Pumping and photon losses through the cavity
mirrors are described with a master equation, which is solved in order to 
determine the stationary density matrix. The photon first-order correlation
function, from which the spectral function is found, is computed with the 
help of the Quantum Regression Theorem. 
The spectral function qualitatively describes the polariton lasing
regime in the model, corresponding to pumping rates two orders of
magnitude lower than those needed for ordinary (photon) lasing. The
second-order coherence functions for the photon and the electron-hole 
subsystems are computed as functions of the pumping rate.
\end{abstract}

\maketitle

\section{Introduction}
\label{sec1}

Excitonic polaritons are quasiparticles made up from strongly coupled 
electron-hole pairs and photons\cite{polaritons1,polaritons2}. They are 
experimentally realized in semiconductor optical microcavities with
embedded quantum wells. The small volume of the microcavity, high
reflectivity of its walls, and quasiresonance condition between the 
confined-photon and excitonic energies guarantee the strong coupling
regime.

At very-low excitation rates, in mean only a single quasiparticle 
lives inside the cavity.
With increasing excitation power, however, an abrupt increase of 
ground-state occupation takes place due to the quasibosonic statistics 
of the polaritons. A threshold behaviour 
of the photoluminescence is observed. This behaviour has been
interpreted as Bose-Einstein condensation of polaritons 
\cite{Yamamoto,LeSiDang}
or as a dynamical effect (polariton lasing)\cite{JBloch}. The latter 
position is motivated by the experimental demonstration that 
thermalization mechanisms are not effective.

In the present paper, we start from the idea of the polariton laser
\cite{Imamoglu}, where pumping provides a reservoir from which the 
low-lying polariton states are populated. Unlike common lasers, no
population inversion is required and the
active medium (the excitons) is strongly interacting with the cavity
photons, forming the quasibosonic polaritons.

The theoretical description of polaritons faces the difficulties inherent
to a many-particle strongly-interacting system working under a 
non-equilibrium pumping regime. Our strategy to tackle this problem
is based upon two simplifications. First, we consider a finite system
\cite{PhysE35(2006)99,nuestraPRL,distortedGibbs}, that is a single 
photon mode, and a 
finite number of single-particle states (ten) for electrons and holes.
Then, the electron-hole-photon many-particle Hamiltonian is numerically
diagonalized in order to find the energies and wavefunctions of the
system. We stress that both Coulomb and electron-hole-photon interactions
are treated exactly in our scheme. Second, we compute the stationary
density matrix from a master equation which accounts for photon losses
through the cavity mirrors and pumping. The master equation is solved in a 
truncated set of many-particle states. Notice that these simplifications
preserve the main ingredients of the problem: the existence of fermionic
and bosonic degrees of freedom, the strong coupling between them, the
existence of a finite number of single-particle states for fermions
(around $10^4$ in Ref. [\onlinecite{JBloch}], 10 in our model)
participating in the conformation of polaritons, a
stationary state reached when pumping and losses are equilibrated, etc.

Strictly speaking, our model describes a quantum dot supporting a few
excitonic states and strongly interacting with the lowest photon mode of
a thin micropillar. It covers an intermediate region between the 
two-level dot \cite{Tejedor1,Tejedor2,Tejedor3} and the infinite system 
(well) \cite{previos}. It is simple enough to allow exact diagonalization 
but, at the same time, complex enough to capture many of the properties 
of the infinite system.

The plan of the paper is as follows. In Sec. \ref{sec2}, the model is described
in details. In the next section, we briefly sketch the algorithm for the
numerical diagonalization of the Hamiltonian, and show a few results for the 
energy spectrum and
matrix elements of operators. In Sec. \ref{sec4}, we present the master equation
for the density matrix and show typical occupations of many-polariton levels for
low, intermediate, and relatively strong pumping rates. In Sec. \ref{sec5},
the way of obtaining the exact photoluminescence (PL) spectral function, and the
approximations leading to the simplified expression used in the paper are
clarified. From this expression, we compute the intensity, position and linewidth
of the main PL peak as functions of the pumping rate. Sec. \ref{sec6} is devoted 
to second-order coherence functions. Finally, in the last section, we summarize 
the main results of the paper.

\section{The model polariton system}
\label{sec2}

\begin{figure}[t]
\begin{center}
\includegraphics[width=.95\linewidth,angle=0]{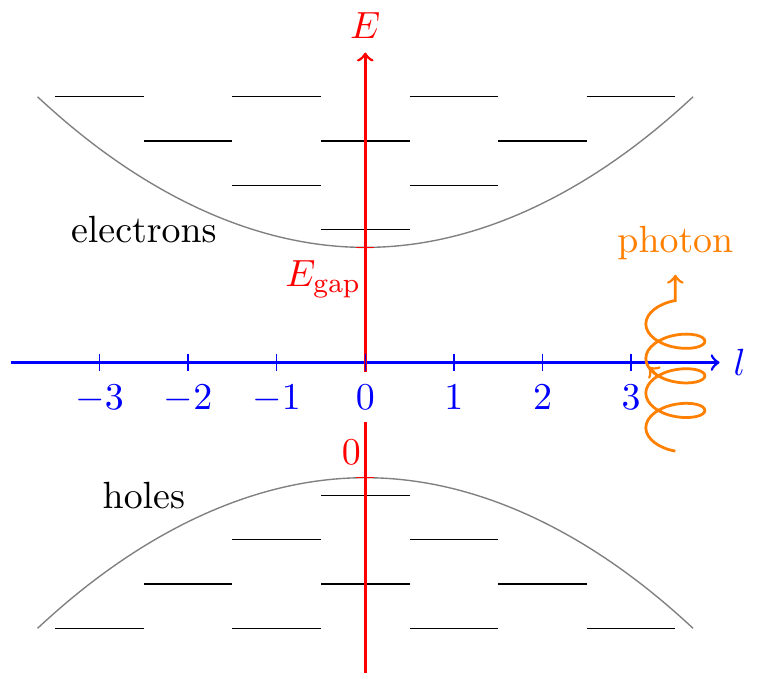}
\caption{\label{fig1n} (Color online) Schematic representation of the 
harmonic oscillator states for electrons and holes in the model.}
\end{center}
\end{figure}

As mentioned in the preceding section, we study a finite polariton system. 
A GaAs micropillar with radius of about one micron or lower is considered, 
in such a way that the lowest photon mode is well separated from the higher 
modes \cite{mp}, and we can assume that a single photon mode is coupled to 
the lowest electron-hole states. The active medium inside the cavity is
described by a finite number of harmonic-oscillator states for electrons and
holes, as shown in Fig. \ref{fig1n}. The number of single-particle states (ten)
is dictated only by practical reasons: the dimension of the many-particle 
Hilbert space grows exponentially with the number of states. This finite
system could be a good model for a quantum dot inside a thin micropillar, and
even could be used to obtain the qualitative behaviour of quantum well-based
micropillars.

The interaction Hamiltonian includes electron-electron, hole-hole, and
electron-hole Coulomb interactions as well as electron-hole-photon coupling,
the latter in the rotating-wave approximation \cite{HK}:

\begin{eqnarray}
H&=& \sum_{i}\left\{T^{(e)}_{i}e^\dagger_i e_i+T^{(h)}_i h^\dagger_i h_i
 \right\} \nonumber\\
&+& \frac{\beta}{2} \sum_{ijrs}\langle i,j||r,s\rangle~ 
 e^\dagger_i e^\dagger_j e_s e_r 
 +\frac{\beta}{2} \sum_{ijrs}\langle i,j||r,s\rangle~ 
 h^\dagger_i h^\dagger_j h_s h_r
\nonumber\\
&-& \beta \sum_{ijrs}\langle i,j||r,s\rangle~ 
 e^\dagger_i h^\dagger_j h_s e_r+(E_{gap}+\Delta)~ a^\dagger a\nonumber\\
&+& g\sum_i\left\{ a^\dagger h_{\bar i}e_i+a e^\dagger_i h^\dagger_{\bar i}
 \right\}.
\label{eq1n}
\end{eqnarray}

The effective band gap, $E_{gap}$, is taken as 1500 meV for GaAs. $\Delta$ is
the detuning of the photon mode with respect to $E_{gap}$. The harmonic
oscillator energies are much smaller than $E_{gap}$. We will neglect them
in the single-particle energies of electrons and holes, and will write:
$T_i^{(h)}=0$, $T_i^{(e)}=E_{gap}$. $g$ is the electron-hole-photon coupling 
strength. Notice that we are including
only spin-up electrons, spin-down holes and one ``circular'' polarization of 
photons in Eq. (\ref{eq1n}). A model with the two photon polarizations, which, 
however, would dramatically increase the 
dimension of the Hilbert space, would make possible the study of interesting 
features such as the spontaneous build-up of coherence between ``left-handed'' and
``right-handed'' polaritons \cite{Kavokin}. $\beta$ is the strength of 
Coulomb interactions, and $\langle i,j||r,s\rangle$ -- the dimensionless matrix 
elements among harmonic oscillator states.

The oscillator states are labeled by two quantum numbers: the number of zeroes
in the radial wave function, $k$, and the angular momentum projection along the 
cavity axis, $l$.  The hole state, $\bar i$, in Eq. (\ref{eq1n}) is the conjugate of
the electron state $i$, that is, has the same $k$, but the momentum is $-l$. This
means that the photon interacts only with electron-hole pairs with zero angular
momentum. As a consequence, the total angular momentum of the electron-hole 
system:

\begin{equation}
L=\sum_i \left(l_i^{(e)}+l_i^{(h)}\right),
\end{equation}

\noindent
is a conserved magnitude. In addition, the Hamiltonian, Eq. (\ref{eq1n}), preserves 
the polariton number: 

\begin{equation}
N_{pol}=N_{pairs}+N_{ph}=\frac{1}{2}\sum_i \left(e_i^{\dagger}e_i
 +h_i^{\dagger}h_i\right) + a^{\dagger}a.
\end{equation}

We notice the similarity between ours and a finite Dicke model \cite{Dicke}. The
infinite Dicke model has been used to describe polaritons in microcavities 
\cite{Littlewood}. The main difference with our approach is the following. In the 
Dicke model of polaritons, we first solve for the excitons and retain only the 
ground state. Multiexcitonic states are not considered. This, may be, is a good 
approximation for far-appart, small (not supporting multiexcitons) quantum dots in 
a microcavity.

\begin{figure}[t]
\begin{center}
\includegraphics[width=.95\linewidth,angle=0]{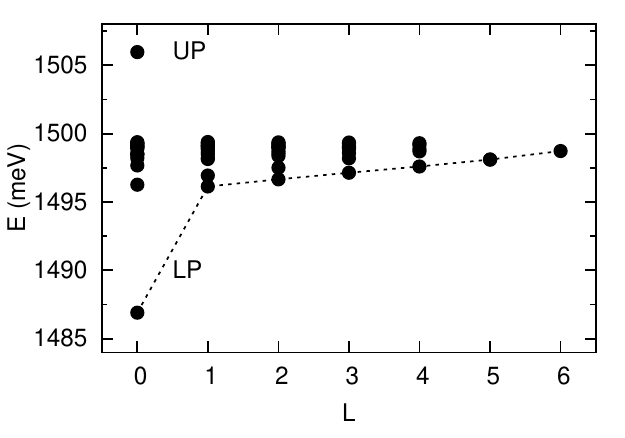}
\caption{\label{fig2n} Polariton states with $N_{pol}=1$.} 
\end{center}
\end{figure}

Many-particle states with fixed $N_{pol}$ and $L$ are constructed in the next
section. We give here a preview in order to compare with the traditional picture
of non-interacting polaritons. We take for the parameters the values, 
$g=3$ meV, and
$\beta=2$ meV. The latter is a reasonable value for GaAs, leading to an
exciton binding energy of a few meVs. The high value of $g$ is, however, 
not intended to be realistic. It is chosen in order to illustrate the 
interesting regime, not studied so far,  where photon-pair coupling and 
Coulomb interactions are comparable. In Fig \ref{fig2n} we show all of 
the states with  $N_{pol}=1$ in the model.  
We joined with a dashed line the lowest states in each 
$L$ tower (the yrast states) in order to conform the lower polariton (LP) branch.
The upper polariton (UP) states, on the other hand, can be identified from the
photoluminescence (PL) emission. We indicated in Fig. \ref{fig2n} the UP state in
the $L=0$ sector. Notice that, because of the strong electron-hole-photon coupling
constant, the UP state is pushed up to high energies in our model. In between LP
and UP states there is a set of ``dark'' polariton states. They play an
important role in the dynamics because they can not decay through photon emission.
Let us stress that, our high-$g$ regime could be of interest in other 
contexts, where ultra-high light-matter couplings have been reported 
\cite{Aji}.

We shall see in Sec. \ref{sec4} that photon losses in the cavity and incoherent
pumping can be modeled by two terms in the master equation for the density 
matrix. We will not include relaxation mechanisms inducing transitions between
states in the same $N_{pol}$ sector (acoustical phonons). As a result, the total
angular momentum is conserved even when pumping and losses are taken into account.
We will solve the dynamics in the $L=0$ tower, which will allow us to compute
the PL emission along the pillar symmetry axis.

Finally, let us comment about the truncation of the basis of single-particle
states in Fig. \ref{fig1n}. For small quantum dots, this is a natural assumption.
In thin micropillars, the number of states strongly coupled to the lowest photon 
mode is large, but finite. In Ref. \onlinecite{JBloch}, for example, it should be
around $10^4$. In this sense, our model may be thought of as a scaled version of 
a micropillar. At larger excitation energies the electron-hole states behave
incoherently and act as a reservoir for the lower polariton states. We partially 
take account of these higher excited states in our model of incoherent pumping
(Sec. \ref{sec4}). Coulomb interactions between polariton states and the
reservoir, which is an additional source of decoherence, will be,
however, neglected.

\section{Exact diagonalization results for the isolated system}
\label{sec3}

For given $N_{pol}$ and $L$, we diagonalize the Hamiltonian in a basis 
constructed from Slater determinants for electrons and holes and Fock states of 
photons. The wave functions are looked for as linear combinations:

\begin{equation}
|I\rangle=\sum C_{S_e,S_h,n} |S_e,S_h,n\rangle.
\end{equation}

\noindent
where $S_e$ and $S_h$ are Slater determinants for electrons and holes 
with the same number of particles, $N_{pairs}$, and the number of 
photons is $n=N_{pol}-N_{pairs}$. When $N_{pol}=0$
there is only one state, the vacuum. When $N_{pol}=1$ there are 17
states with $L=0$. One of them is the state with one photon (no
pairs), and the remaining 16 states correspond to matter excitations 
(no photons), that is, all possible combinations of one electron and one 
hole states with total angular momentum equal to zero. On the other
hand, there are 256 states with $N_{pol}=2$, 1746 states with 
$N_{pol}=3$, etc. As $N_{pol}$ 
increases, the number of eigenstates of $H$ rises, reaching around 18000 
for $N_{pol}\ge 10$. We use Lanczos algorithms \cite{Lanczos} to obtain the 
energies and wavefunctions of the lowest states in each sector.

\begin{figure}[t]
\begin{center}
\includegraphics[width=.95\linewidth,angle=0]{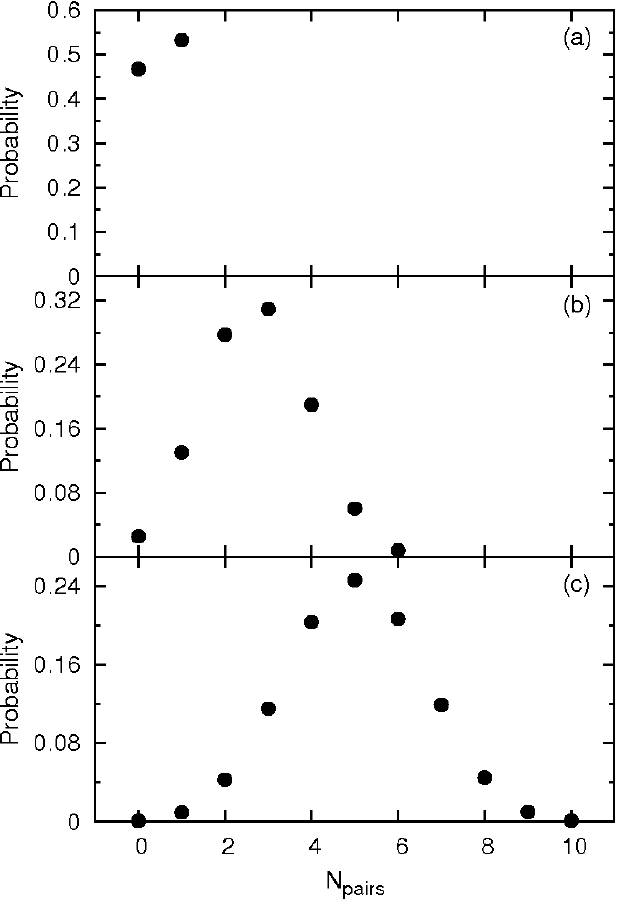}
\caption{\label{fig3n} Weights of the sectors with given $N_{pairs}$ in 
the ground state wave functions. Case (a) corresponds to 
$L=0$ and $N_{pol}=1$, case (b) to $N_{pol}=6$, and case (c) to 
$N_{pol}=600$.} 
\end{center}
\end{figure}

We give in Fig. \ref{fig3n} a schematic representation of the ground state 
wave functions with quantum number $L=0$, and polariton numbers $N_{pol}=1$ 
(case (a)), $N_{pol}=6$ (case (b)), and $N_{pol}=600$ (case (c)). The
detuning parameter is fixed to $\Delta=-3$ meV. This value corresponds to
quasi resonance. Indeed, in the $N_{pol}=6$ case, the distribution is peaked
around $N_{pairs}=3$, whereas in the large-$N_{pol}$ limit it is peaked
around $N_{pairs}=5$, that is the mean occupation of fermionic levels
is near 1/2. Notice that the mean number of photons is around 595 in
the latter case.

\begin{figure}[t]
\begin{center}
\includegraphics[width=.95\linewidth,angle=0]{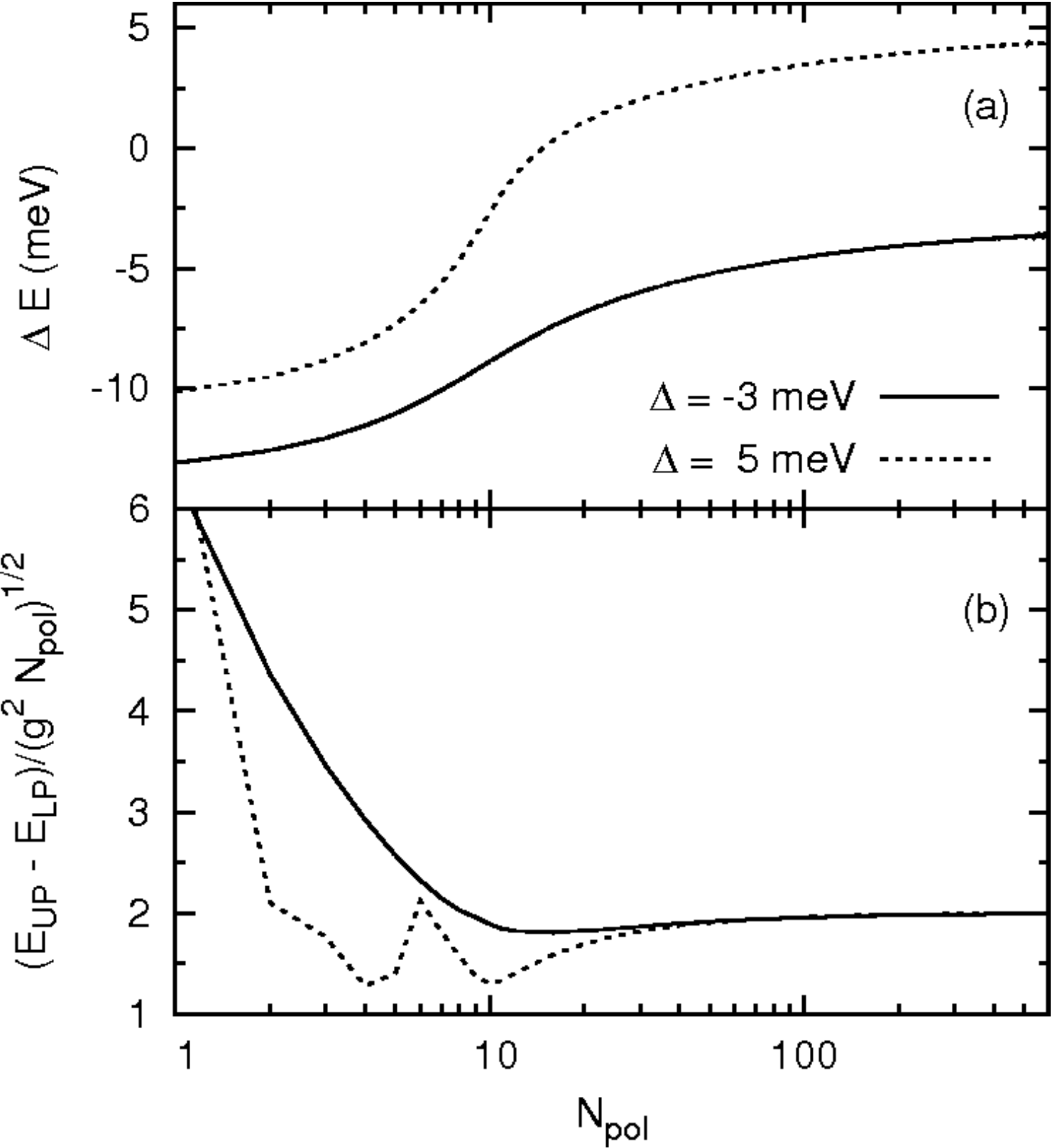}
\caption{\label{fig4n} (a) Energy shift of photon emission from the 
ground state as a function of $N_{pol}$. That is, $\Delta
E=E_{gs}(N_{pol})-E_{gs}(N_{pol}-1)-E_{gap}$.
(b) Scaling in the $E_{UP}-E_{LP}$ energy difference.} 
\end{center}
\end{figure}

In Fig. \ref{fig4n} (a) the many-particle effects on polariton (photon) emission 
are made evident. We plotted the energy difference $E_{gs}(N_{pol})-
E_{gs}(N_{pol}-1)-E_{gap}$ as a function of $N_{pol}$. A persistent
blueshift towards the photon energy (equal to $\Delta$) is noticed as
$N_{pol}$ is increased.

On the other hand, in Fig. \ref{fig4n} (b) the energy difference $E_{UP}-E_{LP}$ 
is plotted as a function of $N_{pol}$. For large
$N_{pol}$ numbers, this difference behaves like $2 \sqrt{g^2 N_{pol}}$.

\begin{figure}[t]
\begin{center}
\includegraphics[width=.95\linewidth,angle=0]{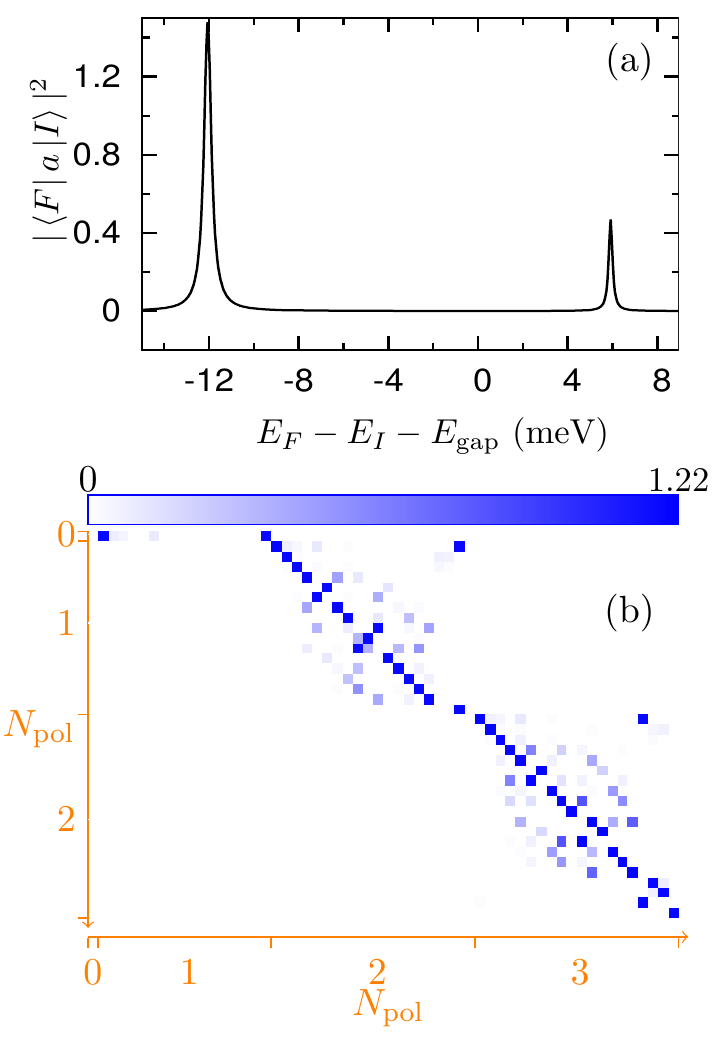}
\caption{\label{fig5n} (Color online) (a) The matrix elements 
$|\langle F |a|I\rangle|^2$ for transitions from $N_{pol}=2$ states to the 
one-polariton ground state. A Lorentzian with $\Gamma=0.1$ meV is used to 
smear out the transitions. (b) The matrix $|\langle F |a|I\rangle|$ in the 
low-$N_{pol}$ sectors.} 
\end{center}
\end{figure}

The obtained wave functions may be used to compute matrix elements of 
operators. As it will be seen in the next section, the most important matrix
elements related to photon emission and losses are $\langle F |a|I\rangle$,
where the many-polariton states $F$ and $I$ are such that $N_{pol}(F)=
N_{pol}(I)-1$. We show in Fig. \ref{fig5n} (a) the matrix elements squared
$|\langle F |a|I\rangle|^2$ for transitions from $N_{pol}=2$ states to the 
one-polariton ground state. A Lorentzian with $\Gamma=0.1$ meV is used to 
smear out the transitions. The analogs of UP and LP states are also
clearly distinguished here and in any $N_{pol}$ sector. The transfer of
population from the UP state with $N_{pol}$ polaritons to the LP state
with $N_{pol}-1$ polaritons will be a key ingredient in the dynamics, 
as will become clear in the next section.

In Fig. \ref{fig5n} (b) we draw the absolute value of the matrix 
elements $|\langle F |a|I\rangle|$ in the 
low-$N_{pol}$ sectors. When $N_{pol} > 1$, only the lowest 20 states 
are used to construct the matrix. Notice that the analogs of LP and UP 
states are
always included among these 20 states. We computed the matrix elements
for $N_{pol}\le 600$ (a matrix of dimension around 12000) and stored 
them in a file. A second file contains the energy eigenvalues. They
are the input files for the dynamics, discussed in the next section.

\section{Master equation description of pumping and losses}
\label{sec4}

The actual polariton system is not isolated. Photons escape mainly
through the cavity mirrors. The spontaneous pair decay through leaky
modes of the cavity is much less important \cite{Tejedor1}, and will be
neglected. In order to maintain a mean number of polaritons in the cavity, the 
system should be continuously pumped. As mentioned before, pumping comes
from excited pair states decoupled from the photon field, which may
decay through emission of optical phonons, for example. We will, 
however, neglect the effects of Coulomb interactions with the excited 
pair states on the relaxation of the polariton
states, and also neglect relaxation due to the emission of acoustical 
phonons by the polariton states, prevented by the bottleneck effect 
\cite{bottleneck}, that is selection rules for energy and momentum of 
the transitions which can not be simultaneously satisfied. These two 
effects, i.e. relaxation due to Coulomb interactions or to phonons, 
could be included in a latter stage, but in the present paper they will 
not be considered. This means that the density matrix of the polariton 
system should be determined from a dynamical equation. 

We will use a quantum dissipative master equation \cite{libroOC,Tejedor1} 
in order to describe photon losses and pumping:

\begin{eqnarray}
\frac{{\rm d}\rho}{{\rm d}t}&=&-\frac{i}{\hbar} [H,\rho]
+\frac{\kappa}{2} (2 a\rho a^{\dagger}-a^{\dagger} a \rho-
\rho a^{\dagger}a)\nonumber\\
&+&\frac{P}{2} \sum_{I,J}(2\sigma_{IJ}^{\dagger}\rho\sigma_{IJ}-
\sigma_{IJ}\sigma_{IJ}^{\dagger}\rho
-\rho\sigma_{IJ}\sigma_{IJ}^{\dagger}), 
\label{eq5n}
\end{eqnarray}

\noindent
The parameter $\kappa$ accounts for photon losses through the cavity 
mirrors ($\hbar\kappa\approx E_{gap}/Q$, where $Q$ is the cavity 
quality factor). In our calculations, we take $\kappa=0.1$ ps$^{-1}$. 
Notice that $\kappa << g/\hbar$, thus our model system works under the 
strong light-matter coupling regime. On the other hand, the parameter 
$P$ is a pumping rate. We will
use a sort of homogeneous pumping, with equal probabilities for all
states. To this end, we introduce lowering and rising operators, 
$\sigma_{IJ}|I\rangle=|J\rangle$, 
$\sigma_{IJ}^{\dagger}|J\rangle=|I\rangle$, where
$N_{pol}(I)=N_{pol}(J)+1$. As we are employing a 
finite number of states (20) in each sector with given $N_{pol}>1$, 
total pumping probabilities are finite. The absence of phonon 
thermalization is also the reason why $L=0$ states are decoupled from 
other states with $L\ne 0$. Thus, we will solve Eqs. (\ref{eq5n}) in
the most relevant $L=0$ sector. In addition, we will focus on the
stationary solutions of Eqs. (\ref{eq6n}), that is, the l.h.s. of these
equations equal to zero. 

The number of variables in Eqs. (\ref{eq5n}) may be estimated as 
follows. In each sector with $N_{pol}>1$ there are 20 occupations,
$\rho_{II}$, and $20\times 19=380$ coherences, $\rho_{FI}$, with
$F\ne I$. That is, 400 variables per sector. If we include sectors
with $0\le N_{pol}<N_{pol}^{(max)}$, the total number of variables is
$400 N_{pol}^{(max)}-2$. When $N_{pol}^{(max)}=10$, for example, the
system has 3998 equations. 

We solve the resulting linear system of equations for the stationary
density matrix with $N_{pol}^{(max)}=10$, and found the remarkable 
fact that the coherences are three order of magnitude lower than the 
occupations, that is the density matrix is approximately diagonal in the
energy representation \cite{distortedGibbs}. For example, for the set
of parameters $\Delta=-3$ meV, $P=0.01$ ps$^{-1}$, we get: 
$\sum_{I\ne J}|\rho_{IJ}|/\sum_I \rho_{II}=7\times 10^{-4}$.

\begin{figure}[t]
\begin{center}
\includegraphics[width=.95\linewidth,angle=0]{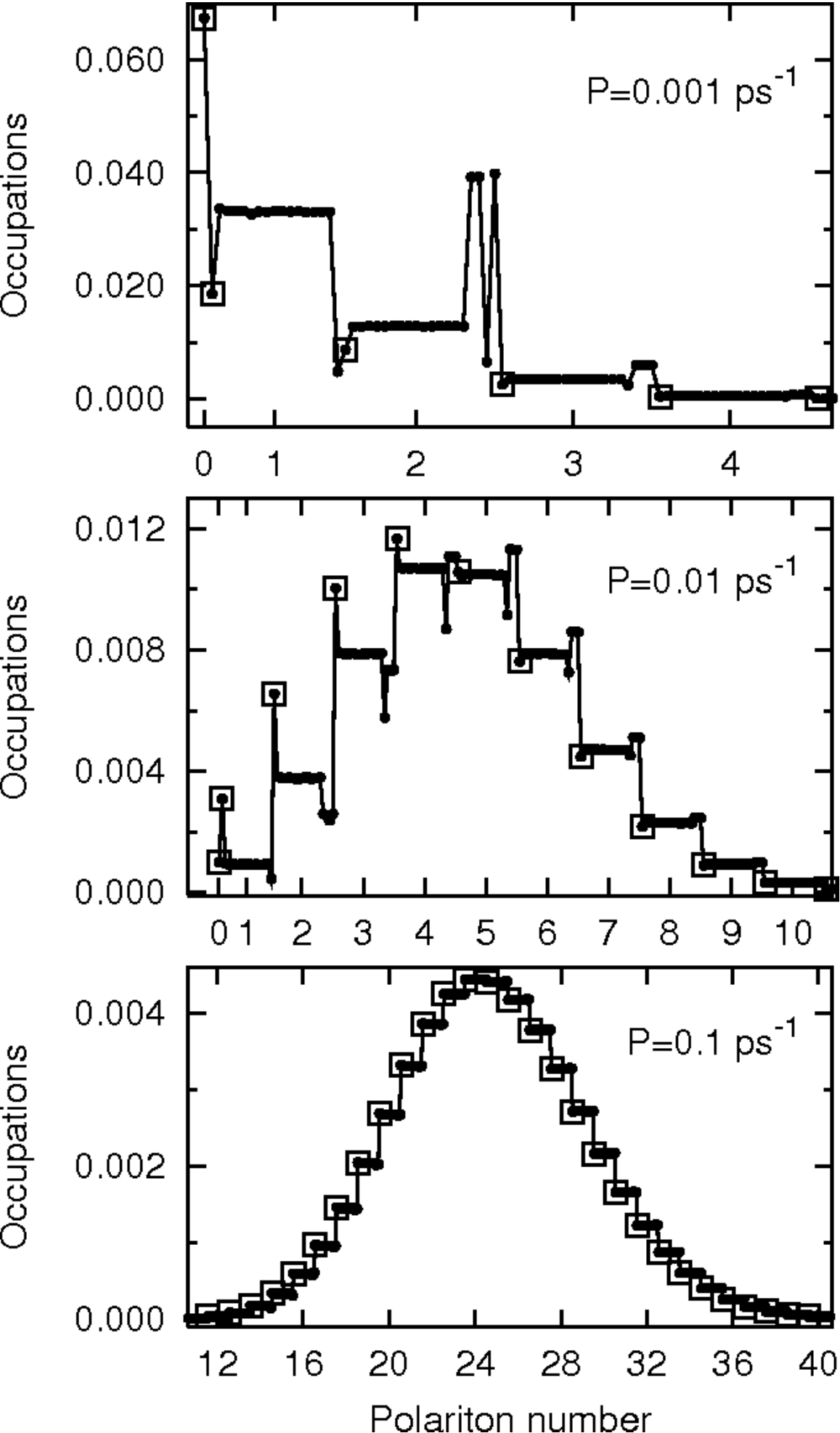}
\caption{\label{fig6n} Occupations at three different pumping rates:
 low (upper panel), intermediate (polariton laser, central panel), and
 high pumping (lower panel). The detuning parameter is $\Delta=-3$ meV.}
\end{center}
\end{figure}

In what follows, in order to extend the analysis up to relatively high 
polariton numbers ($N_{pol}^{(max)}=600$), we will neglect the
coherences. The number of variables reduces to $20 N_{pol}^{(max)}-2$.
For the occupations in the stationary limit, Eqs. (\ref{eq5n}) take 
the explicit form:

\begin{eqnarray}
0&=& \kappa \sum_{J} |\langle I|a|J\rangle|^2\rho_{JJ} 
-\kappa\rho_{II}\sum_J|\langle J|a|I\rangle|^2
\nonumber\\
&+& P\sum_{N_{pol}(J)=N_{pol}(I)-1}\rho_{JJ}- P\rho_{II}N_{up}(I), 
\label{eq6n}
\end{eqnarray}

\noindent
where $N_{up}(I)$ counts the number of states with polariton
number $N_{pol}(I)+1$.  We have $N_{up}(1)=17$, 
$N_{up}(I)=20$ for $1\le N_{pol}(I)< N_{pol}^{(max)}$ and, finally,
$N_{up}(I)=0$ for $N_{pol}(I)=N_{pol}^{(max)}$.

The set of homogeneous linear equations (\ref{eq6n}) should be
complemented with the constraint, 

\begin{equation}
\sum_I \rho_{II}=1,
\end{equation}

\noindent
which corresponds to the conservation of probability.

We show in Fig. \ref{fig6n} three regimes of pumping: low, 
intermediate, and large pumping rates, clearly differentiated by 
the patterns of occupations. In that
figure, the $y$-axis corresponds to the occupations $\rho_{II}$, whereas
in the $x$-axis the states are arranged in increasing order of the 
polariton number, $N_{pol}$. Recall that the first state is the vacuum
with $N_{pol}=0$, then we have 17 states with $N_{pol}=1$, then 20
states with $N_{pol}=2$, etc. The ground state in each sector with
fixed $N_{pol}$ is indicated by a square.

At low pumping rates, the mean polariton number, defined as
$\langle N_{pol}\rangle=\sum_I \rho_{II}N_{pol}(I)$, is 
$\langle N_{pol}\rangle\approx 1$. The state with the highest occupation
is the vacumm. The ground-state occupations in sectors with $N_{pol}>1$ are
depressed. On the other hand, in the situation represented in the 
central panel of Fig. \ref{fig6n}, $\langle N_{pol}\rangle$ is around
four. The ground state occupations in sectors with 
$N_{pol}<\langle N_{pol}\rangle$ are enhanced with respect to the other
states in the same sector. This is a kind of stimulated
occupation of ground states. Finally, for large pumping rates the
occupation in each sector with fixed $N_{pol}$ is nearly uniform. In the
example shown in the lower panel of Fig. \ref{fig6n}, 
$\langle N_{pol}\rangle$ is around 24. A broad bell of occupied states
ranging from $N_{pol}\approx 12$ to around 40 is observed. 

Once computed the stationary density matrix, one can estimate the
photoluminescence response in the stationary state.

\section{The photoluminescence spectral function}
\label{sec5}

In order to obtain the photoluminescence spectral function, $S(\omega)$,
we follow the lines sketched in paper [\onlinecite{Tejedor1}]. 
$S(\omega)$ is defined in terms of the first-order correlation function 
of photons:

\begin{equation}
S(\omega)=\frac{1}{\pi} {\rm Re} \int_0^{\infty}{\rm d}\tau
 \exp (-i\omega\tau) \langle a^{\dagger}(t+\tau) a(t)\rangle.
\end{equation}

\noindent
This function is to be computed with the help of the Quantum Regression
Theorem \cite{libroOC}, which states that if we write:

\begin{equation}
\langle a^{\dagger}(t+\tau) a(t)\rangle=\sum_{I,J} 
 \langle J|a^{\dagger}|I\rangle~g_{a,IJ},
\end{equation}

\noindent
the auxiliary operator:

\begin{equation}
 g_{a,IJ}=\langle |J\rangle\langle I|(t+\tau)~a(t)\rangle,
\end{equation}

\noindent
satisfies with respect to $\tau$ the same master equation as the 
matrix elements $\rho_{IJ}$, with initial conditions:

\begin{equation}
 \left. g_{a,IJ}\right|_{\tau=0}=\sum_K \langle I|a|K\rangle ~
  \rho_{KJ}(t).
\end{equation}

\noindent
In the stationary limit, $t\to\infty$, we get $\rho_{KJ}(t)=
\rho_{JJ}^{(\infty)}\delta_{KJ}$, and $g_{a,IJ}(\tau\to 0)=
\langle I|a|J\rangle~\rho_{JJ}^{(\infty)}$. These initial conditions
dictate that $g_{a,IJ}$ behaves in the same way as the 
``vertical'' coherences, that is, $N_{pol}(I)=N_{pol}(J)-1$. Recall the 
equation for the vertical coherences, which may be obtained from Eq.
(\ref{eq5n}):

\begin{eqnarray}
\frac{\rm d}{{\rm d}\tau} g_{a,IJ}&=&(i\omega_{IJ}-\Gamma_{IJ})
 g_{a,IJ}\nonumber\\
 &+& \kappa\sum_{K,M}\langle I|a|M\rangle g_{a,MK}
  \langle K|a^{\dagger}|J\rangle\nonumber\\
 &-&\frac{\kappa}{2}\sum_{K\ne I,M}\langle I|a^{\dagger}|M\rangle
  \langle M|a|K\rangle g_{a,KJ}\nonumber\\ 
 &-&\frac{\kappa}{2}\sum_{K,M\ne J}g_{a,IM}
 \langle M|a^{\dagger}|K\rangle\langle K|a|J\rangle,
 \label{eq7}
\end{eqnarray}

\noindent
where $\omega_{IJ}=(E_J-E_I)/\hbar$, and

\begin{eqnarray}
\Gamma_{IJ}&=&\frac{\kappa}{2}\sum_K\left\{ \right|\langle K|a|I\rangle|^2
 +|\langle K|a|J\rangle|^2\}\nonumber\\
 &+&\frac{P}{2}\{N_{up}(I)+N_{up}(J)\}.
\end{eqnarray}

The general solution of Eq. (\ref{eq7}) is

\begin{equation}
 g_{a,IJ}=\sum_n C_n e^{\lambda_n\tau} X_{IJ}^{(n)},
\end{equation}

\noindent
where $\lambda_n$ and $X_{IJ}^{(n)}$ are, respectively, the eigenvalues
and eigenvectors of matrix $B_{IJ,MK}$, defined by the r.h.s. of
Eq. (\ref{eq7}), that is

\begin{equation}
 \sum_{M,K} B_{IJ,MK} X_{MK}^{(n)}=\lambda_n X_{IJ}^{(n)}.
 \label{eq15}
\end{equation}

\noindent
The coefficients $C_n$ are determined from the initial conditions:

\begin{equation}
 \sum_{n} C_n X_{IJ}^{(n)}=\langle I|a|J\rangle \rho_{JJ}^{(\infty)}.
 \label{eq16}
\end{equation}

The explicit expression for $S(\omega)$ is the following:

\begin{equation}
S(\omega)=-\frac{1}{\pi} \sum_{I,J}\sum_n\frac
{D^{(r)}_{IJ,n}\lambda_n^{(r)}+D^{(i)}_{IJ,n}(\lambda_n^{(i)}-\omega)}
{(\lambda_n^{(r)})^2+(\lambda_n^{(i)}-\omega)^2}.
\label{eq17}
\end{equation}

\noindent
Where $D_{IJ,n}=\langle J|a^{\dagger}|I\rangle C_n X_{IJ}^{(n)}$, and
the supraindexes $r$ and $i$ refer, respectively, to the real and 
imaginary parts of the magnitudes. The dimension of the matrix problems 
given by Eqs. (\ref{eq15}) and (\ref{eq16}) is 
$17+20\times 17+20\times 20\times(N_{pol}^{(max)}-2)$. When 
$N_{pol}^{(max)}=10$, for example, the dimension is 3557.

\begin{figure}[t]
\begin{center}
\includegraphics[width=.95\linewidth,angle=0]{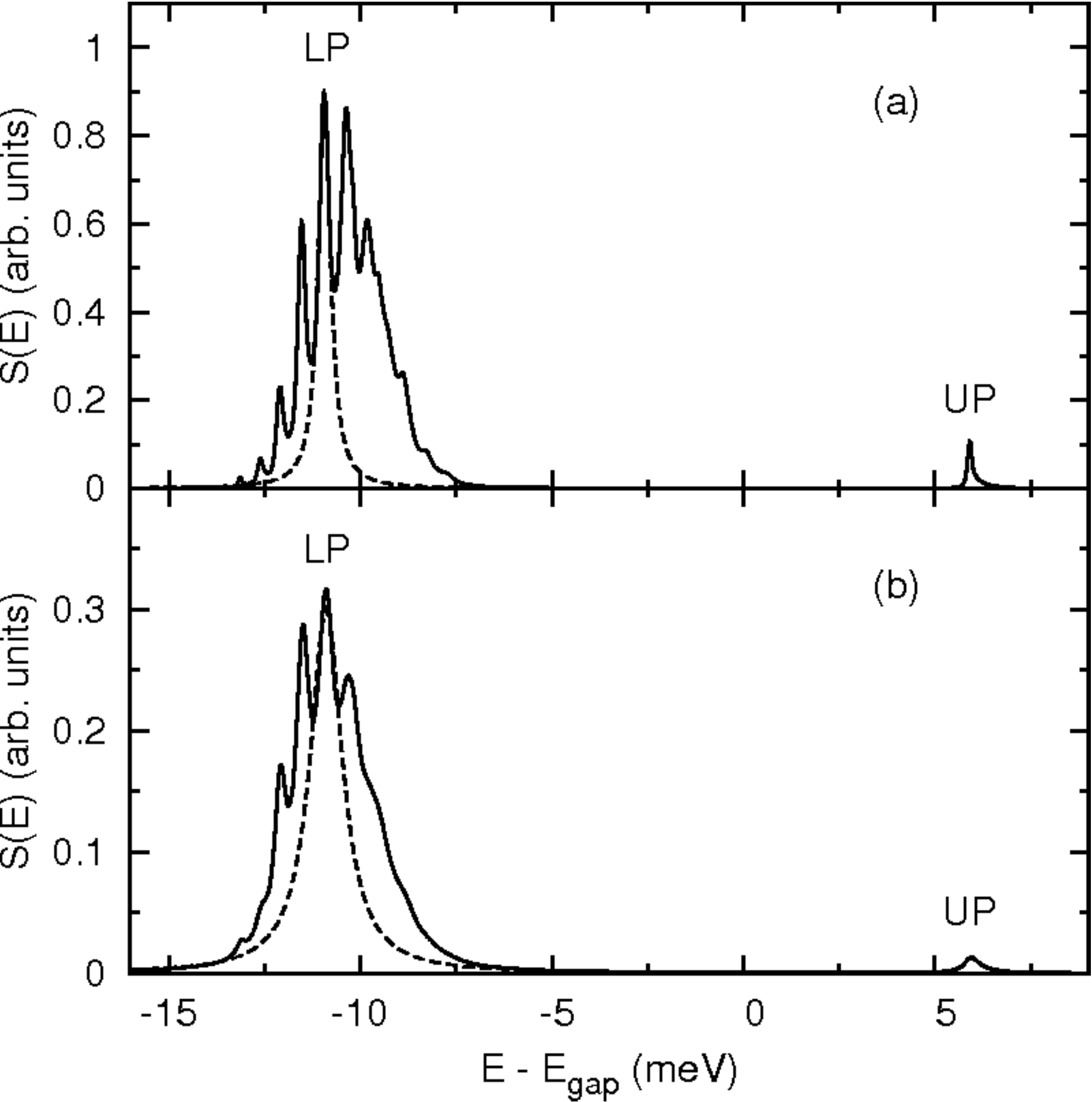}
\caption{\label{fig7n} (a) PL spectral function computed from the exact
expression, Eq. (\ref{eq17}). (b) The spectral function computed from
the approximate expression, Eq. (\ref{eq19}). The detuning is 
$\Delta=-3$ meV, and the pumping rate, $P=0.006$ ps$^{-1}$.}
\end{center}
\end{figure}

We notice that there is an approximate expression for $S(\omega)$ which
is based on the fact that $E_J-E_I\approx E_{gap}\approx 1500$ meV
(for GaAs), whereas $\hbar\kappa$ and $\hbar P$ are smaller than 1 meV. 
In a first approximation, we take only the diagonal terms in 
Eq. (\ref{eq7}), arriving to the following expression for the 
correlation function:

\begin{eqnarray}
&&\left.\langle a^{\dagger}(t+\tau) a(t)\rangle\right|_{t\to\infty}
\nonumber\\
 &&\approx\sum_{I,J}|\langle I|a|J\rangle|^2 ~\rho_{JJ}^{(\infty)}\exp 
 (i\omega_{IJ}-\Gamma_{IJ})\tau,
\end{eqnarray}

\noindent
from which it follows that

\begin{equation}
S(\omega)\approx\frac{1}{\pi}\sum_{I,J}\frac{|\langle I|a|J\rangle|^2
 \rho_{JJ}^{(\infty)}\Gamma_{IJ}}{\Gamma_{IJ}^2+(\omega_{IJ}-\omega)^2}.
 \label{eq19}
\end{equation}

\begin{figure}[t]
\begin{center}
\includegraphics[width=.95\linewidth,angle=0]{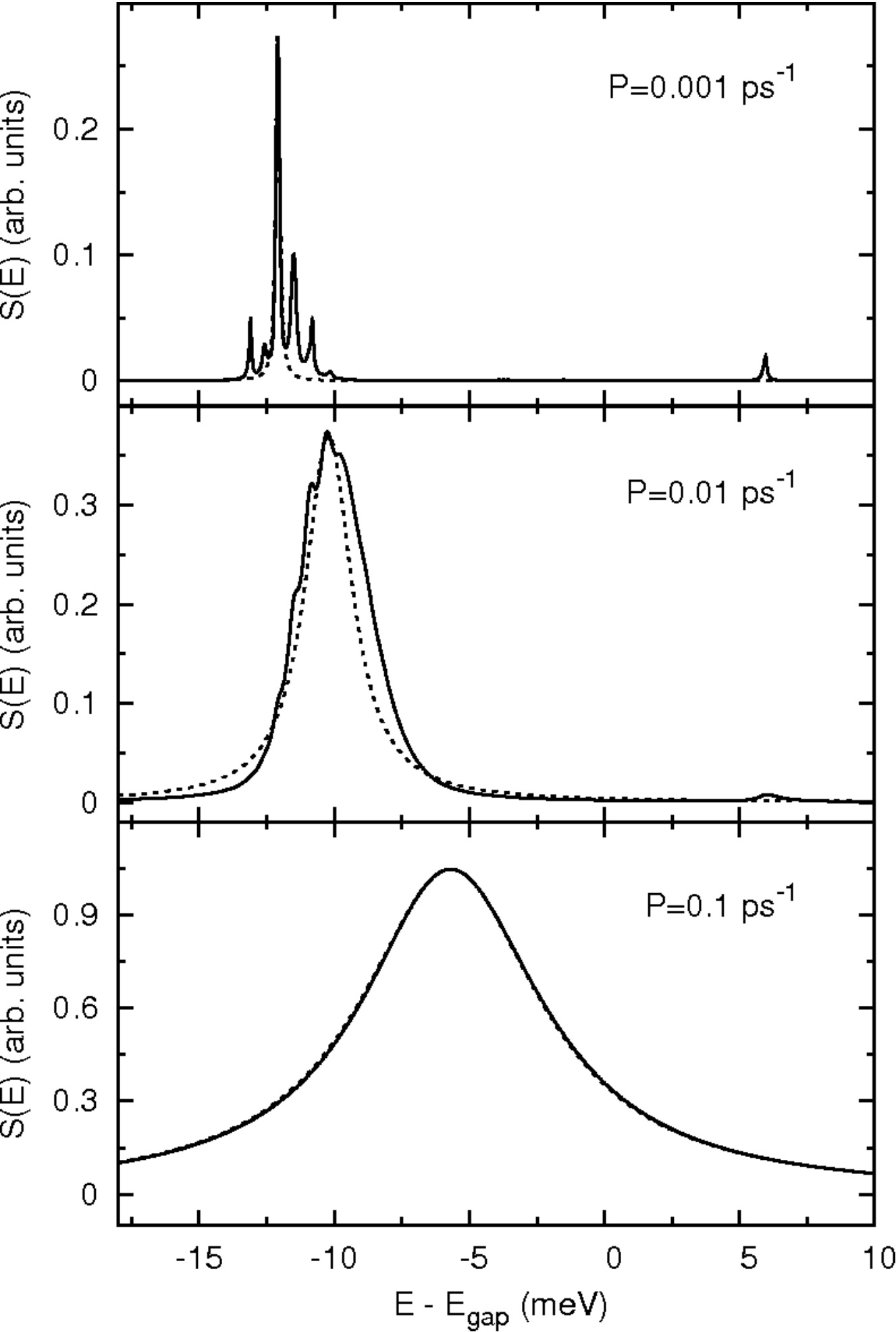}
\caption{\label{fig8n} $S(E)$ and the Lorentzian fit (dashed line) to 
the LP peak for the three cases of Fig. \ref{fig6n}.}
\end{center}
\end{figure}

The main advantage of expression (\ref{eq19}) is the simplicity. The 
luminescence from state $J$ depends on the probability, $\rho_{JJ}$, 
that the state is occupied, and on the matrix elements 
$\langle I|a|J\rangle$ for emission of a photon. The widths 
$\Gamma_{IJ}$ have contributions from losses and pumping, the latter 
is also a source of decoherence. 

\begin{figure}[t]
\begin{center}
\includegraphics[width=.95\linewidth,angle=0]{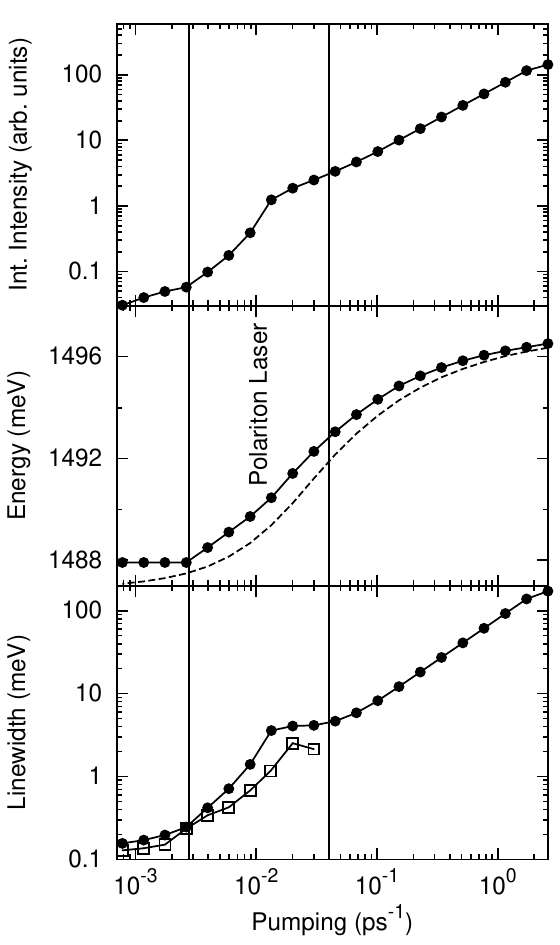}
\caption{\label{fig9n} Integrated intensity, position and linewidth of
 the lower polariton peak, coming from Eq. (\ref{eq19}), as functions of
 the pumping rate. $\Delta=-3$ meV. For the additional lines in the 
 center and bottom panels see explanation in the main text.}
\end{center}
\end{figure}

The nondiagonal terms in Eq. (\ref{eq7}) can only slightly modify the
position of resonances, given by $\omega_{IJ}$. They have a more
appreciable effect on the widths. In Fig. \ref{fig7n}, a comparison is
made between the exact, Eq. (\ref{eq17}), and approximate, Eq. 
(\ref{eq19}), spectral functions. The parameters are such that the mean
number of polaritons is $\langle N_{pol}\rangle=3.4$. The lower energy
emission has contributions from different peaks. We notice, by the way, 
that multimode emission in the polariton lasing regime, which 
is a manifestation of its non-equilibrium nature, has been nicely
demonstrated recently \cite{multimode}. The strongest peak,
which we take as the definition of the LP, is more sharper in the 
exact scheme. We will, nevertheless, use expression (\ref{eq19}) in 
order to obtain the behaviour of the PL even for very strong pumping 
rates ($\langle N_{pol}\rangle\approx 500$), where an effective weak 
coupling regime is established.

\begin{figure}[t]
\begin{center}
\includegraphics[width=.95\linewidth,angle=0]{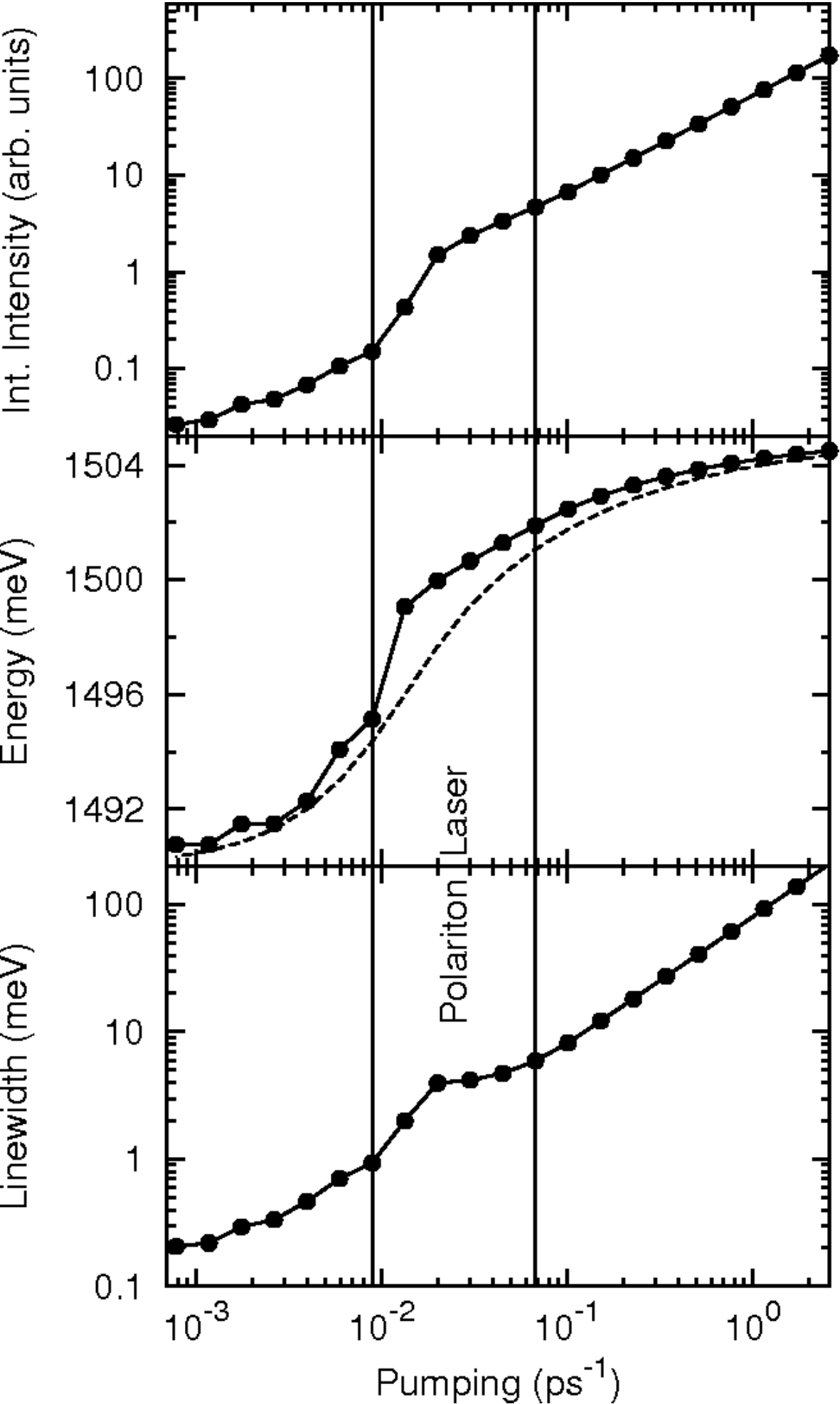}
\caption{\label{fig10n} Same as Fig \ref{fig9n} for $\Delta=+5$ meV.}
\end{center}
\end{figure}

We fit the lower polariton peak to a Lorentzian (dashed line), from 
which the integrated intensity, peak position, and effective linewidth 
are extracted. In
Fig. \ref{fig8n}, we show $S(E)$ and the corresponding Lorentzian fit
for the three cases illustrated in Fig. \ref{fig6n}. The main
characteristics of the polariton emission are apparent in the figure.
That is, a blueshift of the emission, and an increase of the linewidth
as the pumping rate is increased.

The upper panel of Fig. \ref{fig9n} shows the integrated intensity as a
function of $P$ for a detuning $\Delta=-3$ meV. A threshold (change in
the slope) at $P\approx 3\times 10^{-3}$ ps$^{-1}$ is
observed, corresponding to stimulated ground-state occupations when the
number of polaritons exceeds one. This is the ``polariton laser''
regime. At this threshold value, the peak
position (center panel) begins a continuous blueshift towards the bare 
photon energy (1500-3=1497 meV), and the linewidth (bottom panel) starts 
increasing. In the ``polariton laser'' regime there is an interval 
where the linewidth saturates, and even decreases. This is better seen 
in the additional line (empty squares), computed from the exact
equations, Eqs. (\ref{eq15}-\ref{eq17}). The decrease of the linewidth
corresponds to maximum coherence, as will become evident below.

\begin{figure}[t]
\begin{center}
\includegraphics[width=.95\linewidth,angle=0]{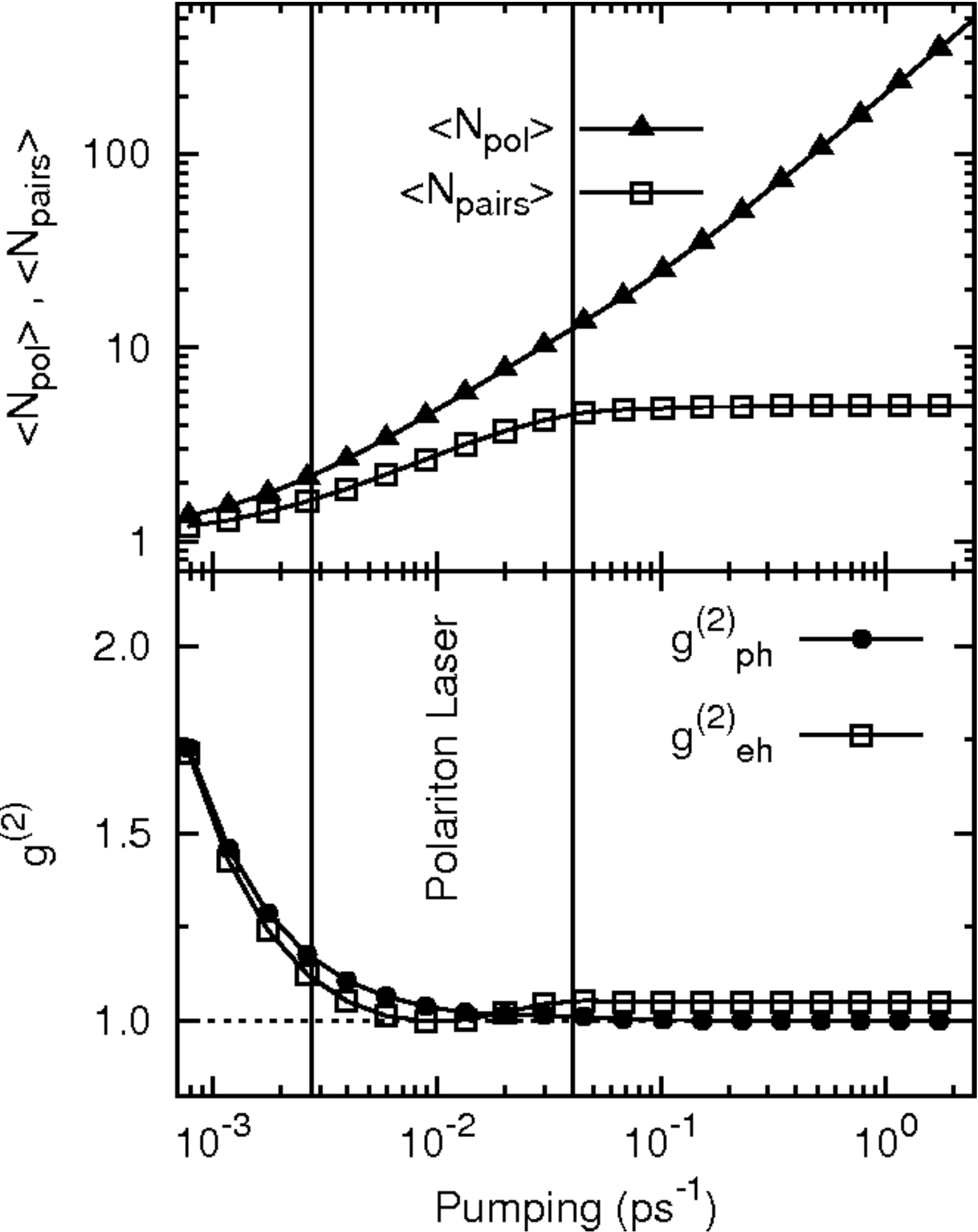}
\caption{\label{fig11n} Upper panel: Mean number of polaritons and 
 electron-hole pairs as a function of $P$. Lower panel: The second-order 
 coherence functions at zero time delay for photons and electron-hole 
 pairs. The detuning parameter is $\Delta=-3$ meV.}
\end{center}
\end{figure}

We draw an additional curve (dashed line) in Fig. \ref{fig9n}, center
panel, which refers to ground-state to ground-state transitions. This 
curve is constructed in the following way. For a given $P$, we find
$\langle N_{pol}\rangle$. Then, the energy of the transition from the
ground state of the system with polariton number equal to 
$\langle N_{pol}\rangle$ to the ground state of the system with
polariton number equal to $\langle N_{pol}\rangle-1$ is found from 
Fig. \ref{fig4n}(a). Comparison with this curve shows that the excited
states, and states with polariton number higher than the mean value
determine the position of the LP peak.

\begin{figure}[t]
\begin{center}
\includegraphics[width=.95\linewidth,angle=0]{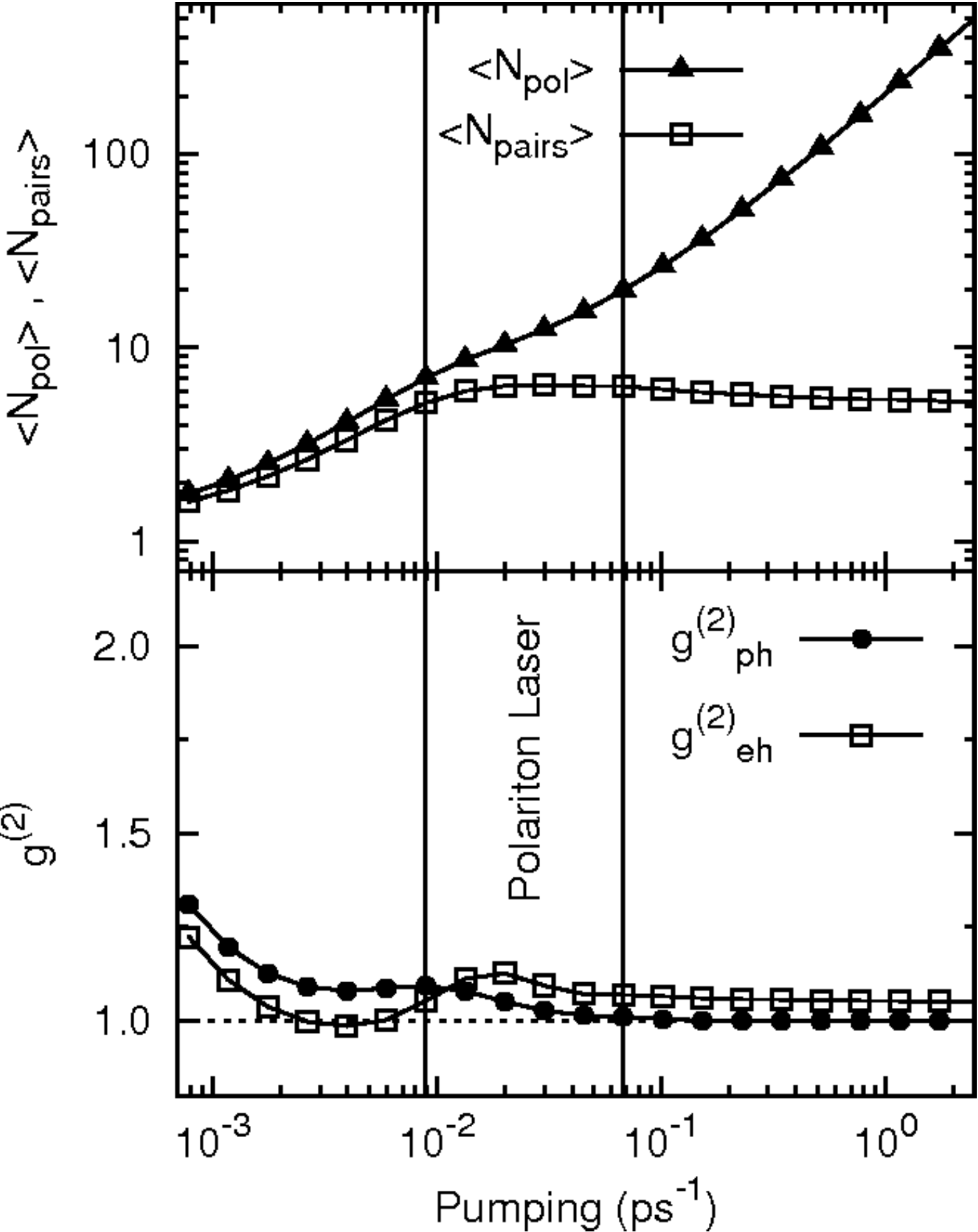}
\caption{\label{fig12n} Same as Fig. \ref{fig11n} for $\Delta=+5$ meV.}
\end{center}
\end{figure}

The right border of the polariton laser regime is conventionally set to
$P\approx 4\times 10^{-2}$ ps$^{-1}$ in this figure. It is 
characterized by a second change of slope in the intensity curve, and 
a renewed increase of linewidth. Let us notice that, in this
quasiresonant case, the mean occupation 
of fermionic levels becomes near one half in the border, a fact that 
could be appreciated below. For strong pumping rates, we observe a 
tendency to saturation in the position of the peak (towards the photon 
energy), which indicates the emergence of a new regime characterized 
by an effective, weak photon-matter interaction. 

In Fig. \ref{fig10n}, we show results for positive detuning,
$\Delta=+5$ meV, where the electron-hole contents of the wave functions 
are higher. At this level, they look very similar to those reported
in Fig. \ref{fig9n}. We can approximately fix the limits of the 
polariton laser regime as $9\times 10^{-3}<P<6\times 10^{-2}$ ps$^{-1}$.
However, as it will become clear in the next section, the mean number
of pairs is close to 5 near threshold, leading to a mean occupation of
fermionic levels, $\langle N_{pairs}\rangle/10$, close to 1/2. This
example shows that polariton lasing is not in antagonism with population
inversion. Or, to set it in a different way, population inversion in
these systems is not synonym of effectively weak pair-photon coupling.
These results could be related to the small number of available states
for fermions or the chosen values of the system
parameters, but anyway they illustrate aspects of principle.

\section{Second-order coherence functions}
\label{sec6}

In Figs. \ref{fig11n} and \ref{fig12n}, we show the mean number of
polaritons, $\langle N_{pol}\rangle$, the mean number of pairs, 
$\langle N_{pairs}\rangle$, and the coherence properties of 
the photon and matter subsystems in the quasi-resonant and in the
positive detuning case, respectively. We define the second-order 
coherence functions for photons and electron-hole pairs in terms of 
the one- and two-point correlation functions at zero time delay:

\begin{equation}
 g^{(2)}_{ph}=\frac{\langle a^{\dagger}a^{\dagger}aa \rangle}
  {\langle a^{\dagger}a\rangle^2}, 
\end{equation}

\begin{equation}
 g^{(2)}_{eh}=\frac{\langle D^{\dagger}D^{\dagger}DD \rangle}
  {\langle D^{\dagger}D\rangle^2}, 
\end{equation}

\noindent
where $a$ is the photon annihilation operator, and 
$D=\sum_i h_{\bar i}e_i$ is the interband dipole operator. 

The coherence
functions evolve from values larger than two at low pumping rates to
values near one (Poisson statistics, perfectly coherent state)
immediately after the threshold. Notice that the electron-hole subsystem
reaches coherence more rapidly than photons ($g^{(2)}_{eh}<g^{(2)}_{ph}$)
possibly because of Coulomb interactions. For large values of $P$, we 
get $g^{(2)}_{eh}>g^{(2)}_{ph}\approx 1$. Notice also that, in the
resonant case, there is a pumping rate for which both $g^{(2)}_{eh}$ 
and $g^{(2)}_{ph}$ are approximately equal to one. This is the point 
of maximum coherence, and  corresponds to a minimum of the linewidth. 

In the positive detuning case, the minimum of the linewidth is reached 
at the point where $g^{(2)}_{eh}$ has a local maximum. For low pumping
rates, $\langle N_{pol}\rangle$ and $\langle N_{pairs}\rangle$ are very
similar. They start differing precisely at the polariton lasing
threshold, where the population of fermions is inverted.

\section{Conclusions}
\label{sec7}

In conclusion, we have computed the stationary density matrix, the
photoluminescence spectral function, and the second-order coherence
functions in a model polariton system describing a multi-level quantum
dot strongly interacting with the lowest photon mode of a microcavity.
The main features of polariton lasing, i.e. blueshift of the emission
peak and increase of the linewidth as the pumping rate is rised, are
reproduced by the model. Unexpected properties, such as the coexistence 
of polariton lasing and population inversion for positive detuning, are 
also manifested.

Our polariton model with a finite number of degrees of freedom, could be
positioned in between the two-level system, studied in Ref.
\onlinecite{Tejedor1}, and the infinite degrees of freedom systems
considered, for example, in Refs. \onlinecite{previos}. Our model is
simple enough to be numerically diagonalized but, at the same time,
complex enough to capture many of the features of the infinite system.
In this sense, our results could be qualitatively 
compared with the experiment reported in Ref. \onlinecite{JBloch},
although the values of our model parameters are completely unrealistic. 

Finally, we should stress that  we are aware of the 
limitations of our model. We understand, for example, that $g^{(2)}$ 
does not rise further in the regime of strong pumping because the model 
does not include higher fermionic levels, which should become populated 
in this regime, that interact with the lowest polariton states, 
partially destroying coherence.

\acknowledgments
This work was supported by the Programa Nacional de Ciencias 
Basicas (Cuba), the Universidad de Antioquia Fund for Research, and 
the Caribbean Network for Quantum Mechanics, Particles and Fields 
(ICTP). Authors are grateful to A. Cabo and P.S.S. Guimaraes for useful 
discussions.

\end{document}